\documentstyle[aps,prl,preprint,floats,epsfig]{revtex}

\begin{document}

\newcommand{\allt}{\mbox{$a_{\ell\ell}^0$}}
\renewcommand{\alt}{\mbox{$a_{\ell}^0$}}
\newcommand{\allm}{\mbox{$a_{\ell\ell}^m$}}
\newcommand{\alm}{\mbox{$a_{\ell}^m$}}
\renewcommand{\ae}{\mbox{$a_{\eta}$}}
\newcommand{\ah}{\mbox{$a_{h}$}}
\newcommand{\ahh}{\mbox{$a_{hh}$}}
\newcommand{\alh}{\mbox{$a_{\ell h}$}}
\newcommand{\af}{\mbox{$a_{f}$}}
\newcommand{\dl}{\mbox{$d_{\ell}$}}
\newcommand{\dll}{\mbox{$d_{\ell\ell}^{like}$}}
\newcommand{\hmu}{\mbox{$h_\mu$}}
\newcommand{\he}{\mbox{$h_e$}}
\newcommand{\hen}{\mbox{$h_{e'}$}}

\preprint{\tighten\vbox{\hbox{\hfil CLNS 01/1717}
                        \hbox{\hfil CLEO 01-01}
}}

\title{\large Bounds on the $CP$ Asymmetry in Like-sign 
Dileptons from $B^{0}\bar{B}^{0}$ Meson Decays}

\author{CLEO Collaboration}
\date{January 5, 2001}

\maketitle
\tighten

\begin{abstract} 
We have measured the charge asymmetry in like-sign dilepton yields from
$B^{0}\bar{B}^{0}$ meson decays using the CLEO detector at the Cornell 
Electron Storage Ring. We find $a_{\ell\ell} \equiv
\left[N(\ell^{+}\ell^{+})-N(\ell^{-}\ell^{-})\right]/
\left[N(\ell^{+}\ell^{+})+N(\ell^{-}\ell^{-})\right] = +0.013 
\pm 0.050 \pm 0.005$. We combine this result with a previous, independent
measurement and obtain $Re(\epsilon_B)/(1 + \vert \epsilon_B \vert^2) =
+0.0035 \pm 0.0103 \pm 0.0015$ (uncertainties are statistical and
systematic, respectively) for the $CP$ impurity parameter,
$\epsilon_B$.
\end{abstract}
\newpage

{
\renewcommand{\thefootnote}{\fnsymbol{footnote}}

\begin{center}
D.~E.~Jaffe,$^{1}$ R.~Mahapatra,$^{1}$ G.~Masek,$^{1}$
H.~P.~Paar,$^{1}$
D.~M.~Asner,$^{2}$ A.~Eppich,$^{2}$ T.~S.~Hill,$^{2}$
R.~J.~Morrison,$^{2}$
R.~A.~Briere,$^{3}$ G.~P.~Chen,$^{3}$
% ...
\end{center}
 
\small
\begin{center}
$^{1}${University of California, San Diego, La Jolla, California 92093}\\
$^{2}${University of California, Santa Barbara, California 93106}\\
$^{3}${Carnegie Mellon University, Pittsburgh, Pennsylvania 15213}\\
% ...
\end{center}

\setcounter{footnote}{0}
}
\newpage

The neutral $B$ mesons mix, just as the neutral kaons do. $K^0 -
\bar{K}^0$ mixing violates $CP$, with different rates for $K^0
\rightarrow \bar{K}^0$ and $\bar{K}^0 \rightarrow K^0$. The Standard
Model predicts that the mixing rates $B^0 \rightarrow \bar{B}^0$ and
$\bar{B}^0 \rightarrow B^0$ are very nearly equal. Thus, a $CP$
asymmetry in $B^0 - \bar{B}^0$ mixing would be evidence for
non-Standard-Model physics.

The mass eigenstates of the neutral $B$ system may be written as
$B_{1,2} = \left[(1 + \epsilon_B) B^0 \pm (1 - \epsilon_B) \bar B^0
\right] /\sqrt{2(1 + \vert \epsilon_B \vert^2)}$, where $\epsilon_B$
is the ``$CP$ impurity parameter'', analogous to the $CP$ violation
parameter $\epsilon$ of $K^{0}$ mixing. If the real part of
$\epsilon_B$ is non-zero, then a $CP$ asymmetry exists. In an
$\Upsilon(4S) \rightarrow B^0 \bar B^0$ event where both $B$ mesons
undergo semileptonic decay, the presence of like-sign dileptons
indicates mixing. A charge asymmetry of such events, $a_{\ell\ell}
\equiv
\left [ N(\ell^{+}\ell^{+}) -
N(\ell^{-}\ell^{-}) \right ] / \left [
N(\ell^{+}\ell^{+}) + N(\ell^{-}\ell^{-}) \right ] $,
indicates a $CP$ violation, related to $Re(\epsilon_B)$ by
$ a_{\ell\ell} =
4 Re(\epsilon_B) / \left ( 1 + \vert \epsilon_B \vert^2 \right )$. For 
a review of the formalism, see Ref.~\cite{edreview}.

The lepton charge asymmetry in $B\bar{B}$ decays with a single charged
lepton, $a_\ell$, also measures the $CP$ violation parameter but with
reduced sensitivity, because $B^+ B^-$ and unmixed $B^0 \bar{B}^0$
events contribute. In particular\cite{edreview},
\begin{equation}
a_{\ell} =
\chi_{d} \left [f_{00}\tau^2_0/(f_{00}\tau^2_0 + f_{+-}\tau^2_\pm) \right ]
 a_{\ell\ell}.
\label{eq.relation}
\end{equation}
Here $f_{00}$ ($f_{+-}$) is the fraction of $\Upsilon(4S)$ decays leading to
$B^0 \bar B^0$ ($B^+ B^-$), $\tau_0$ ($\tau_\pm$) is the lifetime of the
neutral (charged) $B$ meson, and
$\chi_d$ is the neutral $B$ mixing parameter, the ratio of mixed
events to mixed plus non-mixed neutral events.

The Standard Model prediction\cite{SM} for $Re(\epsilon_B)$ is 
$\sim$10$^{-3}$, while superweak models have predictions\cite{superweak} up
to an order of magnitude larger.  Previous searches by
us\cite{oldmeas,andy}, and by others\cite{CDF,OPAL-1,OPAL-2,ALEPH} have
found no evidence for $CP$ violation within a statistical accuracy
ranging from $\pm$0.07 to $\pm$0.01 in $Re(\epsilon_B)$.

In this Letter we report a measurement of dilepton asymmetry, using a
new technique and ten times more data than our previous dilepton
measurement\cite{oldmeas}.  With this increased statistical accuracy
we reduce systematic errors by combining single lepton asymmetries
with dilepton asymmetries.  This technique renders our systematic errors
negligible and is appropriate for $B$-factory-sized data
samples of hundreds of inverse femtobarns.

The data used in this analysis were taken with the CLEO detector at
the Cornell Electron Storage Ring (CESR), a symmetric $e^+ e^-$
collider.  Our sample consists of 9.1 ${\rm fb}^{-1}$ on the
$\Upsilon$(4S) resonance, and 4.4 ${\rm fb}^{-1}$ at a center-of-mass
energy $\sim$60 MeV below the resonance.  The on-resonance sample
contains 10 million $B \bar B$ events and 30 million continuum events,
while the off-resonance sample contains 15 million continuum events.

The CLEO detector\cite{nim} measures charged particle momenta over
95\% of $4\pi$ steradians with a system of cylindrical drift chambers
immersed in a 1.5~T solenoidal magnetic field. For 2/3 of the data
used here, the innermost tracking chamber was a 3-layer silicon vertex
detector\cite{svx}.  The CLEO barrel and endcap CsI electromagnetic
calorimeters cover 98\% of 4$\pi$. Charged particle species are
identified by specific ionization measurements ($dE/dX$) in the
outermost drift chamber and by time-of-flight counters placed
just beyond the tracking volume.

Muons are identified by their ability to penetrate the iron return
yoke of the magnet (at least five interaction lengths of material in
this analysis).  Electrons are identified by shower energy to momentum
ratio ($E/P$), track-cluster matching, $dE/dX$, and shower shape.  For
angles relative to the beam line of less than $45^\circ$, electrons
pass through the thick end plates of the drift chamber, and the
quality of electron identification degrades.  We make a distinction
between ``central electrons'', with $\vert \cos \theta \vert \leq
0.7$, and ``non-central electrons'', with $\vert \cos \theta \vert > 0.7$.

In this analysis, we wish to count single leptons and lepton pairs,
with all leptons coming from the primary semileptonic decay of $B$
mesons. There are backgrounds from secondary decays $b \rightarrow c
\rightarrow s \ell \nu$, from $B \rightarrow \psi \rightarrow \ell^+
\ell^-$, from pair-converted photons, from hadrons misidentified as
leptons, and from continuum events.  To reduce these backgrounds, we
do the following: require that the leptons have high momentum, 1.6 --
2.4 GeV/$c$; veto leptons that form a $J/\psi$ or $\psi^\prime$
candidate with any other loosely-identified, same-flavor,
opposite-charge lepton in the event; and veto electrons that appear to
originate from photon conversions.  The momentum requirement
eliminates our sensitivity to leptons from taus involved in
semileptonic $B$ decays. In counting lepton pairs, we allow at most
one lepton to be an electron from the non-central region.  To suppress
continuum events in lepton pairs, we require that the leptons be
non-collinear, with the angle $\theta_{\ell \ell}$ between them
satisfying $-0.8 < \cos \theta_{\ell \ell} < +0.9$ (the 0.9 limit
eliminates a rare tracking error where two nearly identical tracks are
found for one particle).  We subtract the remaining continuum
contribution with our off-resonance data.

From off-resonance-subtracted like-sign dilepton yields, $N^m(\ell^\pm
\ell^\pm)$, we calculate the measured charge asymmetry $a^m_{\ell\ell}
\equiv \left [ N^m(\ell^{+}\ell^{+}) - N^m(\ell^{-}\ell^{-}) \right ]
/ \left [ N^m(\ell^{+}\ell^{+}) + N^m(\ell^{-}\ell^{-}) \right ] $,
which is related to the desired, corrected asymmetry $a^0_{\ell \ell}$
by\footnote{A note on notation: a superscript $m$ indicates a measured
quantity while superscript $0$ indicates a true quantity. For example,
$a^m_{\ell \ell}$ is the measured like sign dilepton asymmetry, which
is the true asymmetry $a^0_{\ell \ell}$ diluted by backgrounds and
possibly biased by false asymmetries. We correct $a^m_{\ell \ell}$ to
obtain $a^0_{\ell \ell}$ as described in the text.}

\begin{equation}
\allm = \frac{\dll \allt + 2\ae + r_1(\alh + \af)}{1+r_1} .
\label{eq.diexpress}
\end{equation}

\noindent Here the dilution $\dll$ is the fraction of like-sign
dilepton pairs that are primary pairs; $\ae$ is the charge asymmetry
in the efficiency for detecting and identifying leptons; $r_1/(1 +
r_1)$ is the fraction of measured like-sign dileptons with one being a
misidentified hadron; $\alh$ is the asymmetry in like-sign
lepton-hadron pairs; and $\af$ is the asymmetry in the probability
that a hadron is misidentified as a lepton.  In
Eq.~\ref{eq.diexpress}, pairs with both tracks being hadrons
misidentified as leptons and terms that are products of asymmetries
are very small compared to the statistical accuracy on $\allm$, and
have been neglected.

We measure the probability that a pion will be misidentified as a
lepton using $\pi^\pm$ tracks from $K^0_S \rightarrow \pi^+ \pi^-$
decays and the probability that a kaon will be misidentified as a
lepton using $K^\pm$ tracks from $D^{*+} \rightarrow \pi^+ D^0
\rightarrow \pi^+ K^+ \pi^-$ (and charge conjugate) decays.  We
combine pion and kaon misidentification probabilities, separately for
positive and negative tracks, using the $K/\pi$ abundance ratio given
by Monte Carlo simulation.  This procedure gives a probability of
0.9\% that a hadron will be misidentified as a muon, 0.04\% as a
central electron, and 0.3\% as a non-central electron.  Using these
numbers with the yields of like-sign lepton-hadron pairs, we find
values of $r_1$ ranging from 0.15 ($\mu \mu$) to 0.07 ($\mu e$) to
0.006 ($e e$).

The charge asymmetries in the misidentification probabilities, $\af$,
are +0.18 $\pm$ 0.05 for muons, --0.50 $\pm$ 1.00 for central
electrons, and +0.36 $\pm$ 0.25 for non-central electrons.  The charge
asymmetries in like-sign lepton-hadron pairs, $\alh$, are small and
have small errors ($\sim$$0.02 \pm 0.02$).  Thus, the correction term
$r_1(\alh + \af )$ in Eq.~\ref{eq.diexpress} contributes very
little to the final uncertainty.  In solving Eq.~\ref{eq.diexpress} for
$a^0_{\ell \ell}$, the term $(1 + r_1)$ multiplies $\allm$.  Since the
error on $\allm$ is comparable to $\allm$ itself, and since $r_1$ is
reasonably well determined, this correction also contributes little to 
the final uncertainty.

Dilution factors $d_{\ell\ell}$ are determined from Monte Carlo
simulation.  For like-sign pairs we find $\dll = 0.70$, while for
opposite-sign pairs $d_{\ell\ell}^{opposite} = 0.96$; for single
leptons, the fraction that are primary is $\dl = 0.97$. 
For example, for like-sign lepton pairs 70\% of events are primary pairs,
22\% are primary-secondary pairs, 7\% are events with one primary
lepton and one lepton from a $J/\psi$ decay, and 2\% are events with a 
primary lepton and the other lepton from a photon conversion.

This leaves all correction terms in Eq.~\ref{eq.diexpress} determined
except $\ae$, the asymmetry in the efficiency for detecting and
identifying leptons, positive {\it vs.} negative.  While this
asymmetry is not expected to be more than 1--2\%, that is
sufficiently large to be important. We see no direct way to measure
$\ae$.  Consequently, we turn to the measured asymmetry in yields for
single leptons, $\alm$.  That asymmetry may be expressed as

\begin{equation}
\alm = \frac{\dl\alt + \ae + r_0(\ah + \af)}{1+r_0}\ .
\label{eq.singexpress}
\end{equation}

\noindent Here $r_0/(1 + r_0)$ is the ratio of the total yield of
misidentified hadrons (total hadron yield times the average
misidentification probability) to the total measured lepton yield,
$\ah$ is the asymmetry in single hadrons, $\alt$ is related to $\allt$
by Eq.~\ref{eq.relation}, and $\af$, $\ae$, and $\dl$ have been
previously defined.  We find $r_0$ equals 0.02 for muons, 0.001 for
central electrons, and 0.01 for non-central electrons.  The value of $\af$
has been previously determined, and $\ah$ is small, with small errors
($\sim$$0.01 \pm 0.01$).  Thus the correction term $r_0(\ah + \af)$
contributes little to the final error.  Similarly, the factor $(1 +
r_0)$ contributes little error.  We are thus able to express $\ae$ in
terms of $\alm$ and $\allt$, and inserting
Eq.~\ref{eq.singexpress} into Eq.~\ref{eq.diexpress}, we obtain

\begin{equation}
\allt = \frac{\allm(1 + r_1) - 2\alm(1 + r_0) - (r_1 - 2r_0)\af}
{\dll - 2 \dl \chi_d \left [f_{00}\tau^2_0/(f_{00}\tau^2_0 + f_{+-}\tau^2_\pm)
\right ]} \ .
\label{eq.combexpress}
\end{equation}

We have outlined our procedure as if there were only one variety of
dilepton pair, while in fact there are five: $\mu \mu$, $\mu e$, $\mu
e^{'}$, $ee$, and $ee^{'}$, where $e$ and $e^{'}$ refer to central and
non-central electron candidates, respectively. Our actual procedure is
to compute a weighted sum of dilepton asymmetries, using
Eq.~\ref{eq.diexpress}, and then eliminate $a^\mu_\eta$, $a^e_\eta$,
and $a^{e^{'}}_\eta$ from it using the three measured single lepton
asymmetries and Eq.~\ref{eq.singexpress}.  Dilepton yields and
asymmetries are given in Table~\ref{tab.di_lep_yields}.  Single lepton
yields and asymmetries are given in Table~\ref{tab.single_lep_yields}.
The combined result is $\allt = +0.013 \pm 0.050$, where the
uncertainty is statistical only.

    From the yields of like-sign and opposite-sign dilepton pairs,
corrected for misidentified hadrons, we calculate the $B^0 \bar B^0$ mixing
parameter $\chi_d$ via
\begin{equation}
\chi_d = \frac{d^{like}_{\ell \ell}(N(\ell^+ \ell^+) + N(\ell^- \ell^-))}
{d^{opposite}_{\ell \ell}N(\ell^+ \ell^-) +
d^{like}_{\ell \ell}(N(\ell^+ \ell^+) + N(\ell^- \ell^-))}
\left(\frac{f_{00}\tau_{0}^{2}+
f_{+-}\tau^{2}_{\pm}}{f_{00}\tau_{0}^{2}}\right)\ .
\label{eq.mixing}
\end{equation}
The dilution-factor-corrected ratio of like-sign to all dilepton pairs 
in Eq.~\ref{eq.mixing} is
consistent among the five varieties of lepton pairs, and averages to
0.081 $\pm$ 0.002.  The term 
$\left(\frac{f_{00}\tau_{0}^{2}+
f_{+-}\tau^{2}_{\pm}}{f_{00}\tau_{0}^{2}}\right)$ corrects the denominator of
Eq.~\ref{eq.mixing} for dilepton pairs from $B^+ B^-$.  We evaluate it using
$f_{+-}\tau_\pm/f_{00}\tau_0 = 1.11 \pm 0.08$\cite{sylvia} and
$\tau_\pm/\tau_0 = 1.06 \pm 0.03$\cite{pdg}, obtaining 0.46 $\pm$ 0.02.  This
gives $\chi_d = 0.175 \pm 0.008$, to be compared with the PDG value\cite{pdg} of
0.174 $\pm$ 0.009.  Note that the value for $\chi_d$ that we obtain depends on
the correctness of $d^{like}_{\ell \ell}$.  Rather than claim a new measurement 
of mixing, we turn things around, and use the good agreement with the PDG value
to place a limit on the error of $d^{like}_{\ell \ell}$ of $\pm$7\% of itself.

    The systematic error of $\pm$7\% of $\dll$ leads to a $\pm$9\%
multiplicative systematic error in $\allt$.  Other multiplicative systematic
errors considered are from $\chi_d$ ($\pm$1.7\%), and from the off-resonance
subtraction ($\pm$1.7\%).  We combine these for an overall multiplicative
systematic error of $\pm$10\%.

    We have considered several additive systematic errors, in
particular the following sources:
imperfect cancellation of $\ae$ between dilepton and single lepton
events due to differences in single and dilepton momentum spectra
($\pm$0.0030); systematic uncertainty in the hadron misidentification
probability ($\pm$0.0037); difference between $\af$ for dileptons and
$\af$ for single leptons (small, included in statistical error);
systematic uncertainty in the off-resonance
subtraction ($\pm$0.0020); and a difference in the momentum scale between
positive and negative tracks ($\pm$0.0006).  These combine to an additive
systematic error of $\pm$0.005.

In conclusion, we have measured the like-sign dilepton charge
asymmetry to be $\allt = (+0.013 \pm 0.050 \pm 0.005)(1.00 \pm 0.10)$,
where the errors are statistical, additive systematic, and
multiplicative systematic, respectively.  This result is more accurate
than our previous dilepton asymmetry measurement\cite{oldmeas}, +$0.03
\pm 0.10 \pm 0.03$, and supplants it.  It is in good agreement with
our recent measurement\cite{andy} of the $B^0$ -- $\bar{B}^0$ mixing
asymmetry via partial hadronic reconstruction, $+0.017 \pm 0.070 \pm
0.014$, and statistically independent of it.  We take a weighted
average of the two measurements, divide the result by 4, and obtain

$$\frac{Re(\epsilon_B)}{1 + \vert \epsilon_B \vert^2} = +0.0035 \pm 0.0103
\pm 0.0015\ .$$

\noindent This result is more accurate than CDF's\cite{CDF} ($+0.025
\pm 0.062 \pm 0.032$, assuming $\epsilon_{B_s} = 0$), of comparable
accuracy to OPAL's\cite{OPAL-1} ($+0.002 \pm 0.007 \pm 0.003$,
assuming $\epsilon_{B_s} = 0$), and a result from ALEPH recently
submitted for publication\cite{ALEPH}. Furthermore, our result is
independent of any assumptions about $B_s$.  It is consistent with
zero, and with the Standard Model predictions, but lacks the
statistical accuracy to see asymmetries as small as those predictions.
The technique of combining dilepton and single lepton asymmetries to
reduce systematic errors will be appropriate for the large data
samples soon to be available at $B$ factories.

We gratefully acknowledge the effort of the CESR staff in providing us with
excellent luminosity and running conditions.
This work was supported by 
the National Science Foundation,
the U.S. Department of Energy,
the Research Corporation,
the Natural Sciences and Engineering Research Council of Canada, 
the Swiss National Science Foundation, 
the Texas Advanced Research Program,
and the Alexander von Humboldt Stiftung.

\begin{table}[h]
\begin{tabular}{c||c|c|c|c}
Sample & $++$ Yield & $--$ Yield & Like-sign Asymmetry & Opposite-sign 
Yield \\ \hline\hline
$\mu \mu$ &   286 $\pm$ 19 & 286 $\pm$ 19 &
$+$0.000 $\pm$ 0.046 & 4395 $\pm$ 78 \\
$e e$ &       205 $\pm$ 17 & 175 $\pm$ 16 &
$+$0.079 $\pm$ 0.062 & 3255 $\pm$ 64 \\
$\mu e$ &     500 $\pm$ 25 & 505 $\pm$ 25 &
$-$0.004 $\pm$ 0.035 & 7713 $\pm$ 92 \\
$\mu e'$ &    163 $\pm$ 16 & 126 $\pm$ 15 &
$+$0.128 $\pm$ 0.078 & 2147 $\pm$ 49 \\
$e e'$ &      103 $\pm$ 19 & 112 $\pm$ 20 & $-$0.042 $\pm$
0.128 & 1797 $\pm$ 59 \\
\end{tabular}
\caption{Yields and asymmetries for dilepton candidates, after
subtraction of scaled off-resonance yields.}
\label{tab.di_lep_yields}
\end{table}

\begin{table}[h]
\begin{tabular}{c||c|c|c}
Sample & $+$ Yield & $-$ Yield & Asymmetry \\ \hline\hline
$\mu$  & 246274 $\pm$ 801 & 246447 $\pm$ 784 &
$-$0.0004 $\pm$ 0.0023 \\
$e$    & 210624 $\pm$ 678 & 208609 $\pm$ 683 &
$+$0.0048 $\pm$ 0.0023 \\
$e'$ & \phantom{0}53766 $\pm$ 435 & \phantom{0}53731 $\pm$ 440 & $+$0.0003 $\pm$
0.0058 \\
\end{tabular}
\caption{Yields and asymmetries for single lepton candidates, after
subtraction of scaled off-resonance yields.}
\label{tab.single_lep_yields}
\end{table}

\end{document}